\documentclass[aps,pra,superscriptaddress,twocolumn]{revtex4}
\usepackage{graphicx}
\usepackage{epstopdf}
\usepackage{tabularx}
\usepackage{amssymb}
\usepackage{amsmath}
\usepackage{bm}
\usepackage{graphpap}
\usepackage{multirow}
\usepackage{notes2bib}
\usepackage{color}

\begin{document}
\title[]{Understanding effects of packing and chemical terminations on the optical excitations of azobenzene-functionalized self-assembled monolayers}
\author{Caterina \surname{Cocchi}}
\email{caterina.cocchi@physik.hu-berlin.de}
\affiliation{Physics Department and IRIS Adlershof, Humboldt-Universit\"at zu Berlin, Berlin, Germany}
\affiliation{European Theoretical Spectroscopic Facility (ETSF)}
\author{Claudia \surname{Draxl}}
\affiliation{Physics Department and IRIS Adlershof, Humboldt-Universit\"at zu Berlin, Berlin, Germany}
\affiliation{European Theoretical Spectroscopic Facility (ETSF)}


\begin{abstract}
In a first-principles study based on many-body perturbation theory, we analyze the optical excitations of azobenzene-functionalized self-assembled monolayers (SAMs) with increasing packing density and different terminations, considering for comparison the corresponding gas-phase molecules and dimers.
The intermolecular coupling increases with the density of the chromophores independently of the functional groups.
The intense $\pi \rightarrow \pi^*$ resonance that triggers photo-isomerization is present in the spectra of isolated dimers and diluted SAMs, but it is almost completely washed out in tightly packed architectures.
Intermolecular coupling is partially inhibited by mixing differently functionalized azobenzene derivatives, in particular when large groups are involved.
In this way, the excitation band inducing the photo-isomerization process is partially preserved and the effects of dense packing partly counterbalanced.
Our results suggest that a tailored design of azobenzene-functionalized SAMs, that optimizes the interplay between the packing density of the chromophores and their termination, can lead to significant improvements in the photo-switching efficiency of these systems.
\end{abstract}
\date{\today}
\maketitle

\section{Introduction}

The goal to exploit the potential of photo-switching molecules in technological applications was hampered for a long time by the challenge of realizing well-ordered functional architectures with high yield~\cite{aben+15nano}. 
Over the years, however, azobenzene-functionalized self-assembled monolayers (SAMs) of alkanethiols have been successfully synthesized~\cite{wolf+95jpc,zhan+97cpl,evan+98lang,yasu+03jacs,delo+05lang,kuma+08nl,weid+08lang,klaj10pac,gnat+15jpcc}.
Unfortunately these systems exhibit the severe drawback of quenched photo-isomerization, due to steric hindrance between the chromophores and their promoted coupling in the excited state~\cite{gahl+10jacs,utec+11pccp}.
To overcome this issue, new experimental strategies have been developed.
Dilution of the azobenzene moieties by selective functionalization of the alkyl chains, as well as modification of the substrate morphology and anchoring of the chromophores to large molecular platforms have considerably improved the switching rate~\cite{jung+10lang,vall+13lang,jaco+14pccp,krek+15lang,mold+15lang,mold+16lang}.
The insertion of small markers covalently bonded to azobenzene has also been exploited to decrease the intermolecular coupling and partially restore the efficient photo-switching behavior of single molecules~\cite{webe+03jpcb,akiy+03jpcb,schm+08apa,naga+09jesrp,wagn+09pccp,gahl+10jacs,bret+12jpcm,masi+14ns}.
With increasing complexity of the SAMs, also the need for a microscopic understanding of the physical mechanisms responsible for the excitation processes in these materials has concomitantly increased.
In a recent study based on many-body perturbation theory, we have shown that local-field effects are primarily responsible for enhancing intermolecular interactions and transition-dipole coupling, thereby significantly reducing the spectral intensity of the absorption band triggering photo-isomerization~\cite{cocc+16jcp}.
We have further demonstrated that the effects of packing density of the chromophores act also in core excitations, by drastically reducing the binding energies of the electron-hole pairs, owing to the interplay between screening, dipole coupling, and wave-function overlap~\cite{cocc-drax15prb}.
Still, a number of questions remains open, in particular regarding the role of functional groups covalently attached to azobenzene.
Indeed, the presence of terminating groups is known to significantly impact the electronic and optical properties of carbon-based systems~\cite{calz+10jpcc,cocc+12jpcc,jin+13nano,li+15jpcc}.
When different functionalizations are mixed in the same SAM sample~\cite{bret+12jpcm}, the system gains additional complexity and it is expected that the interplay with the packing density gives rise to peculiar behaviors.

We aim at addressing these issues in a first-principles work based on a state-of-the-art \textit{ab initio} methodology that explicitly accounts for electron-electron and electron-hole interactions. 
First, we clarify how specific functionalizations of the azobenzene molecules affect the optical excitations upon increasing packing density.
We compare the results obtained for azobenzene-functionalized SAMs with their gas-phase counterparts consisting of single molecules and dimers.
By analyzing the absorption spectra and the character of the electron-hole pairs at decreasing intermolecular distances, we show that the essential spectral features are determined by the properties of the azobenzene backbone and therefore independent of the terminations.
A thorough analysis of the SAMs in relation with the dimer counterparts sheds light on the effects of intermolecular interactions and dipole coupling in the presence of different functionalizations.

\section{Systems}

\begin{figure}
\center
\includegraphics[width=.5\textwidth]{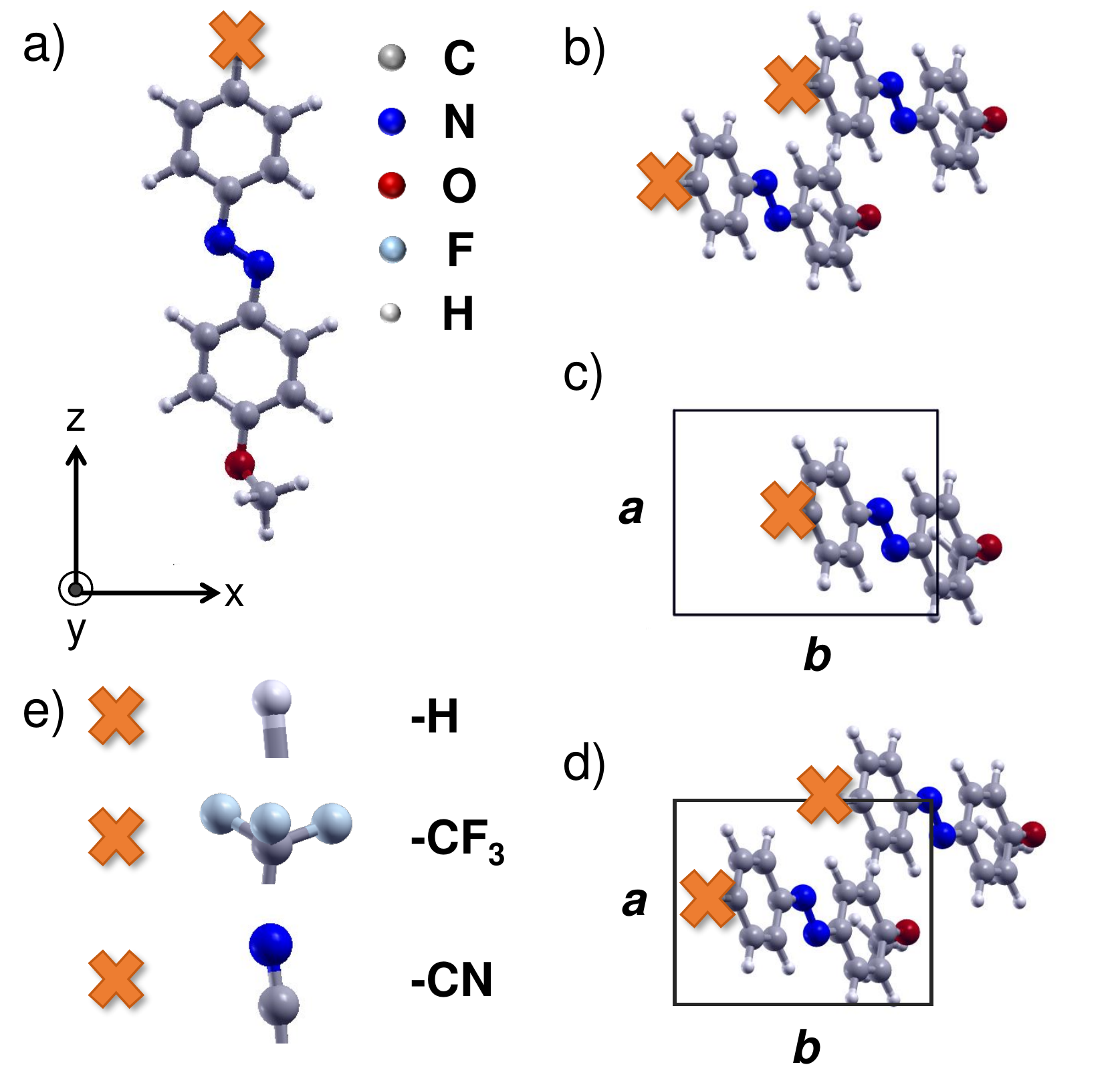}
\caption{(Color online) Ball-and-stick representations of the systems considered in this work: (a) gas-phase molecule, (b) dimer, (c) diluted SAM, and (d) packed SAM. Functionalization marked by crosses, with the corresponding functional groups in panel (e).}
\label{fig:structures}
\end{figure}

We consider here four systems, as sketched in Fig.~\ref{fig:structures}.
The fundamental building block is an azobenzene derivative (Fig.~\ref{fig:structures}a), terminated on one side with a methoxy ($-$OCH$_3$) group to reproduce the chemical environment of the molecule that is covalently attached to the SAM of alkyl chains~\cite{gahl+10jacs,bret+12jpcm,mold+15lang}.
On the opposite end (marked by a cross) the chromophore is functionalized either with a trifluoromethyl ($-$CF$_3$) or with a cyano ($-$CN) group, which are used as markers in experimental samples~\cite{gahl+10jacs,naga+09jesrp,bret+12jpcm,gahl+16ss}. 
We refer to these systems as CF$_3$-az and CN-az, respectively.
H-passivated molecules are also considered in the context of mixed functionalizations.
The effects of packing densities in H-terminated systems have been extensively addressed in a previous publication~\cite{cocc+16jcp}.

Azobenzene-functionalized SAMs are simulated in periodically repeated unit cells modeled according to atomic-force and scanning tunneling microscopy measurements~\cite{wolf+95jpc,jasc+96jpc,mann+02jpcb}.
Densely-packed SAMs (indicated in the following as p-SAMs) include two molecules in an orthorhombic unit cell of lattice vectors $a$=6.05 \AA{} and $b$=7.80 \AA{} (Fig.~\ref{fig:structures}d), such that the minimal distance between the chromophores is less than 4 \AA{}.
In absence of clear experimental indications about the mutual orientation of the two chromophores in the unit cell, we have considered them to be oriented parallel to each other.
Considering that the transition-dipole moments of the azobenzene derivatives are oriented along the long molecular axis~\cite{utec+11pccp}, any rotation around that axis is not expected to affect the optical excitations~\cite{kash+65pac}. 
Only sizable changes to the average tilting angle of the molecules with respect to the substrate could have a significant impact on the spectra.
This scenario is, however, not observed experimentally at high molecular concentrations~\cite{mold+15lang}.
Diluted SAMs (d-SAMs) are modeled in the same unit cell as their packed counterparts, with only one molecule included (Fig.~\ref{fig:structures}c).
As discussed in Ref.~\cite{cocc+16jcp}, these structures represent quite a drastic approximation of the experimental samples, considering the flexible nature of SAMs and the tendency of the alkyl chains to form islands with different chromophore density.
Nonetheless, this choice allows us to rationalize and understand the basic physical mechanisms of optical excitations in such complex materials at increasing packing density of the chromophores.
For comparison, we consider gas-phase dimers (Fig~\ref{fig:structures}b) where the two molecules retain the same mutual distance and orientation adopted in the unit cell of the p-SAM~\cite{utec+11pccp}.
With this, we can clarify whether the optical excitations of p-SAMs can be reproduced by taking into account the interaction of a molecule with only one nearest neighbor.
This procedure is typically adopted in quantum-chemistry calculations, where the standard computational tools do not implement periodic boundary conditions. 
Larger clusters were adopted in more recent works, leading to an improved agreement with experiments~\cite{bena-corn13jpcc,cant+16jpcl,tito+16jpcl}.

To address the effects of different terminations of the azobenzene derivatives we focus on molecules, dimers, and SAMs with one type of termination, as well as on dimers and p-SAMs with mixed functionalizations.
In the former case, both molecules in the unit cell depicted in Fig.~\ref{fig:structures} have the same termination.
In the latter, each of the two chromophores is bonded to a different functional group.
We consider all three possible combinations arising by mixing the groups depicted in Fig.~\ref{fig:structures}, namely CF$_3$-H, CN-H, and CF$_3$-CN dimers and the corresponding p-SAMs.
Azobenzene-functionalized SAMs including a mixing of CF$_3$- and CN-functionalized molecules have been synthesized and characterized in a recent experimental work~\cite{bret+12jpcm}.
\section{Theoretical Background and Computational Details}

Ground-state properties are computed by means of density-functional theory (DFT) \cite{hohe-kohn64pr}, solving the Kohn-Sham (KS) equations \cite{kohn-sham65pr} in the framework of the linearized augmented planewave plus local orbital (LAPW+lo) method.
The electronic structure is calculated from many-body perturbation theory (MBPT) within the $GW$ approximation \cite{hedi65pr}, through the single-shot perturbative approach $G_0W_0$ \cite{hybe-loui85prl}. 
The quasi-particle (QP) correction to the gap is applied as a rigid shift to the KS conduction bands, as a starting point for the solution of the Bethe-Salpeter equation (BSE)~\cite{hank-sham80prb,stri88rnc}, which is adopted to compute optical excitations and absorption spectra, including excitonic effects.
This method has proven to be successful in describing optical excitations of molecular materials, from the gas-phase to the solid state~\cite{ruin+02prl,bussi+02apl,humm+04prl,humm-ambr05prb,humm+05pssb,hahn+05prl,ambr+06cp,herm+08prl,herm-schw11prl,cocc-drax15prb1}.
Within the Tamm-Dancoff approximation the BSE reads in matrix form:
\begin{equation}
\sum_{v'c'\mathbf{k'}} \hat{H}^{BSE}_{vc\mathbf{k},v'c'\mathbf{k'}} A^{\lambda}_{v'c'\mathbf{k'}} = E^{\lambda} A^{\lambda}_{vc\mathbf{k}} ,
\label{eq:BSE}
\end{equation}
where $v$ and $c$ label valence and conduction bands, respectively.
In spin-unpolarized systems, like those considered in this work, the effective two-particle Hamiltonian~\cite{rohl-loui00prb,pusc-ambr02prb} takes the following form:
\begin{equation}
\hat{H}^{BSE} = \hat{H}^{diag} + 2 \hat{H}^x + \hat{H}^{dir}.
\label{eq:H_BSE}
\end{equation}
The \textit{diagonal} term $\hat{H}^{diag}$ accounts for single quasi-particle transitions, while the other two describe the electron-hole ($e$-$h$) interaction.
Specifically, the exchange term
\begin{equation}
\hat{H}^x = \int d^3\mathbf{r} \int d^3\mathbf{r}' \phi_{v\mathbf{k}} (\mathbf{r}) \phi^*_{c\mathbf{k}} (\mathbf{r}) \bar{v}(\mathbf{r},\mathbf{r}') \phi^*_{v'\mathbf{k}'} (\mathbf{r}') \phi_{c'\mathbf{k}'} (\mathbf{r}'),
\label{eq:Hx}
\end{equation}
includes the short-range part of the Coulomb interaction $\bar{v}= v - v_0$, where $v_0 \equiv v(\mathbf{G} = 0)$ is the long-range term subtracted from the total potential.
Following the formalism presented in Refs.~\cite{ambe-kohn60pr,hank78advp}, this choice is allowed when calculating the optical response in terms of the \textit{reducible} polarization function, which behaves as the \textit{irreducible} one in the optical limit ($|\mathbf{q}| \rightarrow 0$)~\cite{kohn58pr}.
In this way, the exchange part of the BSE Hamiltonian accounts only for local-field effects (LFE).
The attractive $e$-$h$ interaction is given by $\hat{H}^{dir}$, which includes the statically screened Coulomb potential $W = \epsilon^{-1} v$,  where $\epsilon$ is calculated within the random-phase approximation:
\begin{equation}
\hat{H}^{dir} \! \! = \! - \! \! \int \! \! \! d^3\mathbf{r} \! \!  \int \! \! d^3\mathbf{r}' \phi_{v\mathbf{k}} (\mathbf{r}) \phi^*_{c\mathbf{k}} (\mathbf{r}') W(\mathbf{r},\mathbf{r}') \phi^*_{v'\mathbf{k}'} (\mathbf{r}) \phi_{c'\mathbf{k}'} (\mathbf{r}').
\label{eq:Hdir}
\end{equation}
Additional information about the formalism and its implementation in the LAPW+lo framework is reported in Refs.~\cite{pusc-ambr02prb,Sagmeister2009}.
The solution of the BSE (Eq. \ref{eq:H_BSE}) yields \textit{singlet} excitations. 
Including only $\hat{H}^{diag}$ corresponds to the independent-particle approximation (IPA).
The resulting spectrum can be used to understand the role of the $e$-$h$ interaction.
For bound excitons, we evaluate binding energies ($E_b$) as the difference between the lowest-energy transition within the IPA, computed including the QP correction, and the excitation energies $E^{\lambda}$ obtained from the solution of the BSE.
Information about character and composition of the excitations is given by the eigenvectors $A^{\lambda}$ of Eq.~\ref{eq:BSE}.
They enter the expression of the exciton wave-function $\Psi^{\lambda}(\mathbf{r}_{e},\mathbf{r}_{h}) = \sum_{v c \mathbf{k}} A^{\lambda}_{vc\mathbf{k}}\phi_{c\mathbf{k}}(\mathbf{r}_{e})\phi^{*}_{v\mathbf{k}}(\mathbf{r}_{h})$, that is a six-dimensional object depending on the hole and the electron coordinates.
This quantity can be visualized in real-space by fixing a point in the hole distribution and plotting the corresponding electron distribution, and viceversa.
A more convenient representation is given in reciprocal space, by means of the \textit{weights} of each vertical transition at a given $\mathbf{k}$-point:
\begin{equation}
w^{\lambda}_{v\mathbf{k}} = \sum_c |A^{\lambda}_{vc\mathbf{k}}|^2 , \, \, \, w^{\lambda}_{c\mathbf{k}} = \sum_v |A^{\lambda}_{vc\mathbf{k}}|^2.
\label{eq:w}
\end{equation}
The sums in Eq.~\eqref{eq:w} run over the range of occupied and unoccupied states considered in the solution of the BSE (Eq. \ref{eq:BSE}).
In this way, the distribution of the hole and of the electron can be visualized simultaneously on top of the band structure.
The eigenvectors $A^{\lambda}$ also enter the expression of the oscillator strength, given by the square modulus of
\begin{equation}
\mathbf{t}^{\lambda}= \sum_{vc\mathbf{k}} A^{\lambda}_{vc\mathbf{k}} \dfrac{\langle v\mathbf{k}|\widehat{\mathbf{p}}|c\mathbf{k}\rangle}{\varepsilon_{c\mathbf{k}} - \varepsilon_{v\mathbf{k}}} ,
\label{eq:t}
\end{equation}
where $\hat{\mathbf{p}}$ is the momentum operator, and $\varepsilon_{v\mathbf{k}}$ and $\varepsilon_{c\mathbf{k}}$ are the QP energies of the involved valence and conduction states, respectively.
To properly account for the non-locality of the electron self-energy, following the arguments proposed in Ref.~\cite{dels-girl93prb} a renormalization term is included in Eq.~\eqref{eq:t}, also for the case where a scissors operator replaces the QP correction.
Optical absorption spectra are expressed by the imaginary part of the macroscopic dielectric tensor:
\begin{equation}
\mathrm{Im}\epsilon_M = \dfrac{8\pi^2}{\Omega} \sum_{\lambda} |\mathbf{t}^{\lambda}|^2 \delta(\omega - E^{\lambda}) ,
\label{eq:ImeM}
\end{equation}
where $\Omega$ is the volume of the unit cell and $\omega$ is the energy of the incident photon.

All calculations are performed using the \texttt{exciting} code~\cite{gula+14jpcm}.
The muffin-tin (MT) radii of C, O, N, F, and H atoms are set to 1.2, 1.2, 1.1, 1.2, and 0.8 bohr, respectively. 
A plane-wave basis-set cutoff $R_{\textrm{MT}}G_{\textrm{max}} = 3.5$ is used.
For the exchange-correlation potential the local-density approximation (LDA) is adopted, employing the Perdew-Wang parameterization~\cite{perd-wang92prb}.
The choice of LDA over the generalized gradient approximation is not expected to affect the electronic properties of the systems (differences in the KS gaps are lower than 0.09 eV), nor the starting point for the $G_0W_0$ calculations~\cite{brun-marq12jctc}.
All calculations for the isolated molecules and dimers are performed in supercells including at least 7.40~\AA{}, 7.65~\AA{}, and 6.35~\AA{} of vacuum in the $x$, $y$, and $z$ direction, respectively, according to the Cartesian axes in Fig.\ref{fig:structures}a, to avoid spurious interactions between the replicas. 
Atomic positions are relaxed until the forces are smaller than 0.025 eV/\AA{}.
Ground-state calculations are performed by sampling the Brillouin zone (BZ) with a $6\times4\times1$ ($3\times2\times1$) \textbf{k}-point grid in case of \textit{packed} (\textit{diluted}) SAMs. 
The same meshes are used also in the $GW$~\cite{nabo+16prb} and BSE calculations~\cite{Sagmeister2009}. 
For computing the screened Coulomb interaction, 200 empty states are employed in $GW$.
For the solution of the BSE, we include 400 empty states in the SAMs and 500 unoccupied orbitals in the isolated molecules.
Local-field effects are accounted for by employing about 500 $|\mathbf{G}+\mathbf{q}|$ vectors in the case of \textit{packed} SAMs, about 10$^3$ for their \textit{diluted} counterparts, and more than 10$^5$ for isolated molecules and dimers.
These parameters ensure convergence within 100 meV of the intense peaks in the BSE spectrum, within the considered optical and near-UV region.
\section{Results and Discussion}
\subsection{Functionalized azobenzene-derived molecules, dimers, and SAMs}
\label{sec:homogeneous}

\begin{figure*}
\center
\includegraphics[width=.8\textwidth]{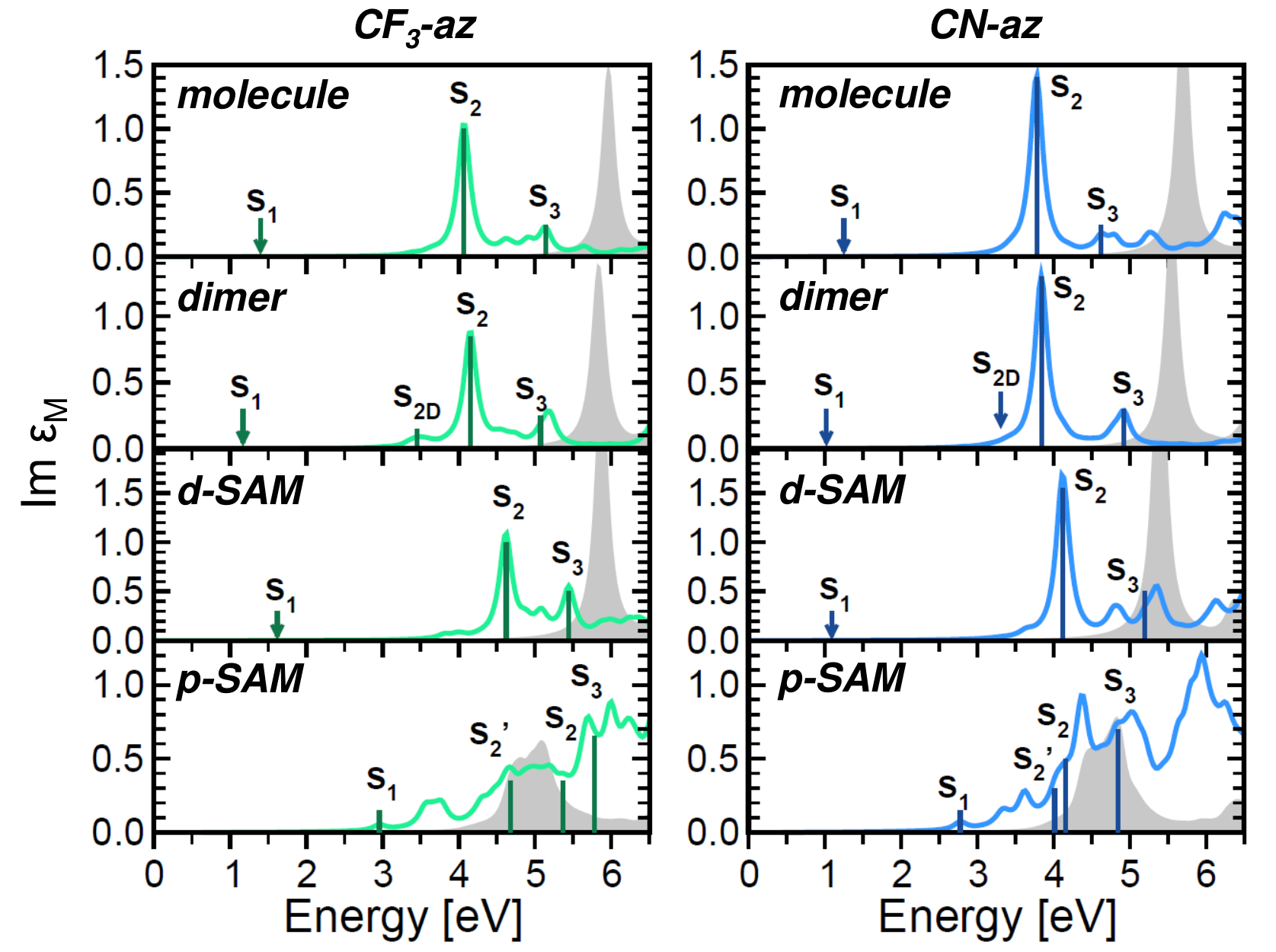}
\caption{(Color online) Optical absorption spectra of the CF$_3$-functionalized (left) and CN-functionalized (right) azobenzene molecule, dimer, and SAMs, diluted (d-SAM) and packed (p-SAM). Im$\epsilon_M$ is averaged over the three Cartesian components. The main excitations are marked by vertical bars of height representative of the oscillator strength. Singlet excitons with vanishing oscillator strength are indicated by arrows. Filled gray areas correspond to the independent-particle approximation (IPA). For the p-SAMs, the IPA spectrum is scaled by a factor 0.25 with respect to the BSE one. A Lorentzian broadening of 100 meV is applied to all spectra to mimic the excitation lifetime.}
\label{fig:homo}
\end{figure*} 

We start our analysis by considering the optical spectra of functionalized azobenzene derivatives, at increasing packing density of the chromophores.
The results are shown in Fig.~\ref{fig:homo} for both $-$CF$_3$ and $-$CN terminations.
In addition to the solution of the BSE (solid lines) we also plot the spectrum obtained by neglecting the $e$-$h$ interaction, \textit{i.e.}, only from vertical transitions including the QP correction (gray area).
The comparison between the two curves is meant to highlight the role of many-body effects.

\begin{table}
\begin{tabular}{cccl}
\hline \hline
\multicolumn{4}{c}{\textit{CF$_3$-az molecule}}\\ \hline 
$\#$ & Excitation energy & Binding energy & Contributions \\
\hline 
S$_1$ & 1.40 & 3.93 & H $\rightarrow$ L (99$\%$)  \\ \hline
S$_2$ & 4.07 & 1.27 & H-1 $\rightarrow$ L (83$\%$)  \\ \hline
\multirow{3}{*}{S$_3$}  & \multirow{3}{*}{5.14} & \multirow{3}{*}{0.20} & H-1 $\rightarrow$ L (29$\%$)  \\ 
& & & H-4 $\rightarrow$ L+3 (21$\%$) \\
& & & H-4 $\rightarrow$ L (14$\%$) \\ 
\hline \hline
\multicolumn{4}{c}{\textit{CF$_3$-az dimer}}\\ 
\hline
$\#$ & Excitation energy & Binding energy & Contributions \\
\hline
\multirow{3}{*}{S$_1$} & \multirow{3}{*}{1.17} & \multirow{3}{*}{4.02} & H $\rightarrow$ L (47$\%$)  \\
& & & H $\rightarrow$ L+1 (41$\%$)  \\
& & & H-1 $\rightarrow$ L+1 (11$\%$)  \\ \hline
\multirow{3}{*}{S$_{2D}$}  & \multirow{3}{*}{3.45} & \multirow{3}{*}{1.74} & H-3 $\rightarrow$ L (18$\%$)  \\ 
& & & H-5 $\rightarrow$ L (28$\%$) \\
& & & H-5 $\rightarrow$ L+1 (18$\%$) \\ \hline
\multirow{2}{*}{S$_2$} & \multirow{2}{*}{4.15} & \multirow{2}{*}{1.04} & H-2 $\rightarrow$ L+1 (34$\%$)  \\
& & & H-3 $\rightarrow$ L (27$\%$)  \\ \hline
S$_3$ & 5.07 & 0.12 & H-2 $\rightarrow$ L+4 (15$\%$) \\
\hline \hline
\multicolumn{4}{c}{\textit{CN-az molecule}}\\ 
\hline
$\#$ & Excitation energy & Binding energy & Contributions \\
\hline
S$_1$ & 1.25 & 3.88 & H $\rightarrow$ L (98$\%$)  \\ \hline
S$_2$ & 3.78 & 1.35 & H-1 $\rightarrow$ L (87$\%$)   \\ \hline
\multirow{4}{*}{S$_3$}  & \multirow{4}{*}{4.62} & \multirow{4}{*}{0.50} & H-1 $\rightarrow$ L+3 (20$\%$)  \\ 
& & & H-1 $\rightarrow$ L+1 (16$\%$) \\
& & & H-3 $\rightarrow$ L (13$\%$) \\ 
& & & H-3 $\rightarrow$ L+5 (13$\%$) \\ \hline
\hline
\multicolumn{4}{c}{\textit{CN-az dimer}}\\ 
\hline
$\#$ & Excitation energy & Binding energy & Contributions \\
\hline
\multirow{2}{*}{S$_1$} & \multirow{2}{*}{1.02} & \multirow{2}{*}{3.90} & H $\rightarrow$ L (50$\%$)  \\
& & & H $\rightarrow$ L+1 (39$\%$)  \\ \hline
\multirow{3}{*}{S$_{2D}$}  & \multirow{3}{*}{3.29} & \multirow{3}{*}{1.63} & H-2 $\rightarrow$ L (48$\%$)  \\ 
& & & H $\rightarrow$ L+2 (13$\%$) \\
& & & H $\rightarrow$ L+3 (12$\%$) \\ \hline
\multirow{2}{*}{S$_2$} & \multirow{2}{*}{3.84} & \multirow{2}{*}{1.09} & H-1 $\rightarrow$ L+3 (47$\%$)  \\
& & & H-3 $\rightarrow$ L (34$\%$)  \\ \hline
\multirow{2}{*}{S$_3$} & \multirow{2}{*}{4.92} & \multirow{2}{*}{--} & H-2 $\rightarrow$ L+3 (15$\%$)  \\
& & & H-3 $\rightarrow$ L+2 (12$\%$)  \\
\hline \hline
\end{tabular}
\caption{Excitation energies, binding energies (for bound excitons only), and dominant single-particle contributions (weights in parenthesis) of the excitations of the CF$_3$- and CN-functionalized azobenzene molecules and dimers marked in Fig.~\ref{fig:homo}. All energies are expressed in eV.}
\label{tab:mol-dimers}
\end{table}

\begin{figure*}
\center
\includegraphics[width=.85\textwidth]{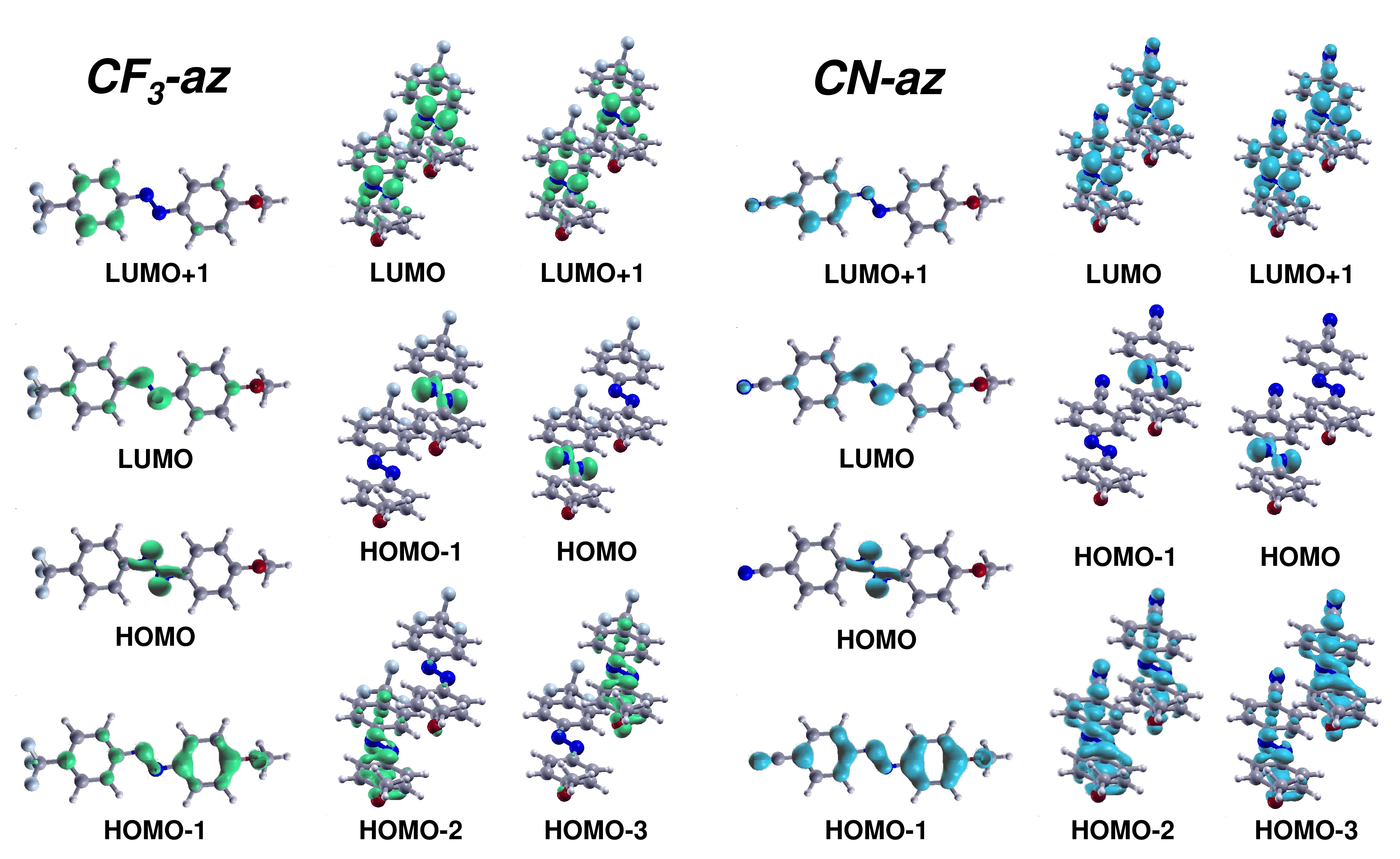}
\caption{(Color online) Highest-occupied and lowest-unoccupied molecular orbitals of the CF$_3$-functionalized (left) and CN-functionalized (right) azobenzene molecule and dimer.}
\label{fig:KS}
\end{figure*}

Regardless of the functionalization, the spectra of the molecules are very similar to each other and retain all the features of the azobenzene backbone~\cite{crec-roit06jpca,cont+08jacs,maur-reut11jcp,cocc+16jcp,fu+17pccp}.
The lowest-energy excitation (S$_1$) is dipole forbidden, stemming from a transition between the non-bonding ($n$) highest-occupied molecular orbital (HOMO) and the lowest-unoccupied one (LUMO), which exhibits $\pi^*$ character (see also Table~\ref{tab:mol-dimers}).
The energy of this excitation is drastically underestimated compared to reference results for azobenzene (see, \textit{e.g.}, Ref.~\cite{crec-roit06jpca}).
This behavior is ascribed to large extent to the starting-point dependence of $G_0W_0$.
Shortcomings of this approach when applied on top of LDA arise for extremely localized states such as the $n$-like HOMO in azobenzene~\cite{brun-marq12jctc}.
The near-UV region is dominated by the strong resonance S$_2$, corresponding to the $\pi$-$\pi^*$ transition between the HOMO-1 and the LUMO, which triggers the photo-isomerization from the \textit{trans} to the \textit{cis} conformation~\cite{rau-lued82jacs}.
In both CF$_3$-az and CN-az molecules this excitation is strongly bound, with $E_b \sim$ 1.3 eV (Table~\ref{tab:mol-dimers}).
The optical excitations of the two molecules are determined by the electronic properties of the \textit{trans}-azobenzene backbone, with negligible influence of the functional groups.
This is confirmed by the character of the molecular orbitals (MOs), plotted for both systems in Fig.~\ref{fig:KS}.
It should be noted though, that in CN-az the distribution of the bonding orbitals HOMO-1, LUMO, and LUMO+1 is not restricted to phenyl rings but spills over to the CN-group.
This slight charge imbalance, however, does not affect significantly the $\pi$/$\pi^*$ nature of the MOs in comparison with pristine azobenzene~\cite{crec-roit06jpca,cont+08jacs,maur-reut11jcp} and in its methoxy-terminated derivative~\cite{cocc+16jcp}.
More pronounced differences arise at higher energies, where we find a number of weak excitations exhibiting again $\pi$-$\pi^*$ character. 
The most intense excitation in this manifold, labeled S$_3$, has binding energy 0.2 eV in CF$_3$-az and 0.5 eV in CN-az.
These different values can be ascribed to the fact that the involved MOs are more affected by the specific termination of the azobenzene derivative, being energetically further away from the gap than those contributing to the S$_2$ resonance (Table~\ref{tab:mol-dimers}).

In the dimers, the overall spectral shape does not change significantly with respect to the gas-phase molecules. 
While the lowest-energy excitation is again dipole-forbidden, stemming from an $n \rightarrow \pi^*$ transition, the intense resonance S$_2$ still dominates the spectrum, at approximately the same energy as in the isolated compound (Fig.~\ref{fig:homo}).
At higher frequencies, we find the S$_3$ excitation, significantly weaker than S$_2$, which is given by a number of $\pi \rightarrow \pi^*$ transitions.
The binding energies of the excitations in the dimers are slightly decreased by approximately 100 meV compared to their counterparts in the single molecules (Table~\ref{tab:mol-dimers}).
This shift comes along with a corresponding reduction of the QP gap, as a consequence of the monomer-monomer interaction.
Moreover, the doubling of the electronic levels in the dimers gives rise to a splitting of the excitonic states.
By examining Fig.~\ref{fig:KS}, we notice that in both CF$_3$-az and CN-az the $n$-like HOMO of the molecule gives rise to two separate levels, with the corresponding wave-functions localized on one single monomer. 
On the other hand, the $\pi$/$\pi^*$ orbitals in the gap region are clearly delocalized on both molecules in the case of CN-az.
In CF$_3$-az this is so for the LUMO and the LUMO+1, while the $\pi$ states HOMO-2 and HOMO-3 tend to segregate on a single chromophore.
This behavior impacts also the optical excitations, which are split into two components, whose intensity is related to the mutual orientation of the molecules in the dimer~\cite{kash+65pac}.
Although this phenomenon, commonly addressed as Davydov splitting~\cite{davy62book}, affects each excitation in the spectrum, here we analyze its effect only on S$_2$.
In the geometry of the dimers, where the molecules are displaced such that the respective transition-dipole moments are almost parallel with respect to each other, the higher-energy component of the excitations is intense while its lower-energy counterpart is almost completely dark~\cite{kash+65pac}.
In the spectra of Fig.~\ref{fig:homo} we indicate the latter excitation S$_{2D}$.
Its composition is analogous to S$_2$, although also additional single-particle transitions contribute to it (see Table~\ref{tab:mol-dimers}), with a corresponding increase of the oscillator strength.
The splitting between S$_{2D}$ and S$_{2}$ amounts to 0.70 eV in the CF$_3$-az dimer and to 0.54 eV in the CN-az one.
These values are overestimated with respect to the results obtained for azobenzene dimers in a recent quantum-chemical study~\cite{tito-saal16jpca}.
Such discrepancies are likely related to the different methodologies adopted and will be systematically analyzed in a dedicated work.

\begin{figure}
\center
\includegraphics[width=.5\textwidth]{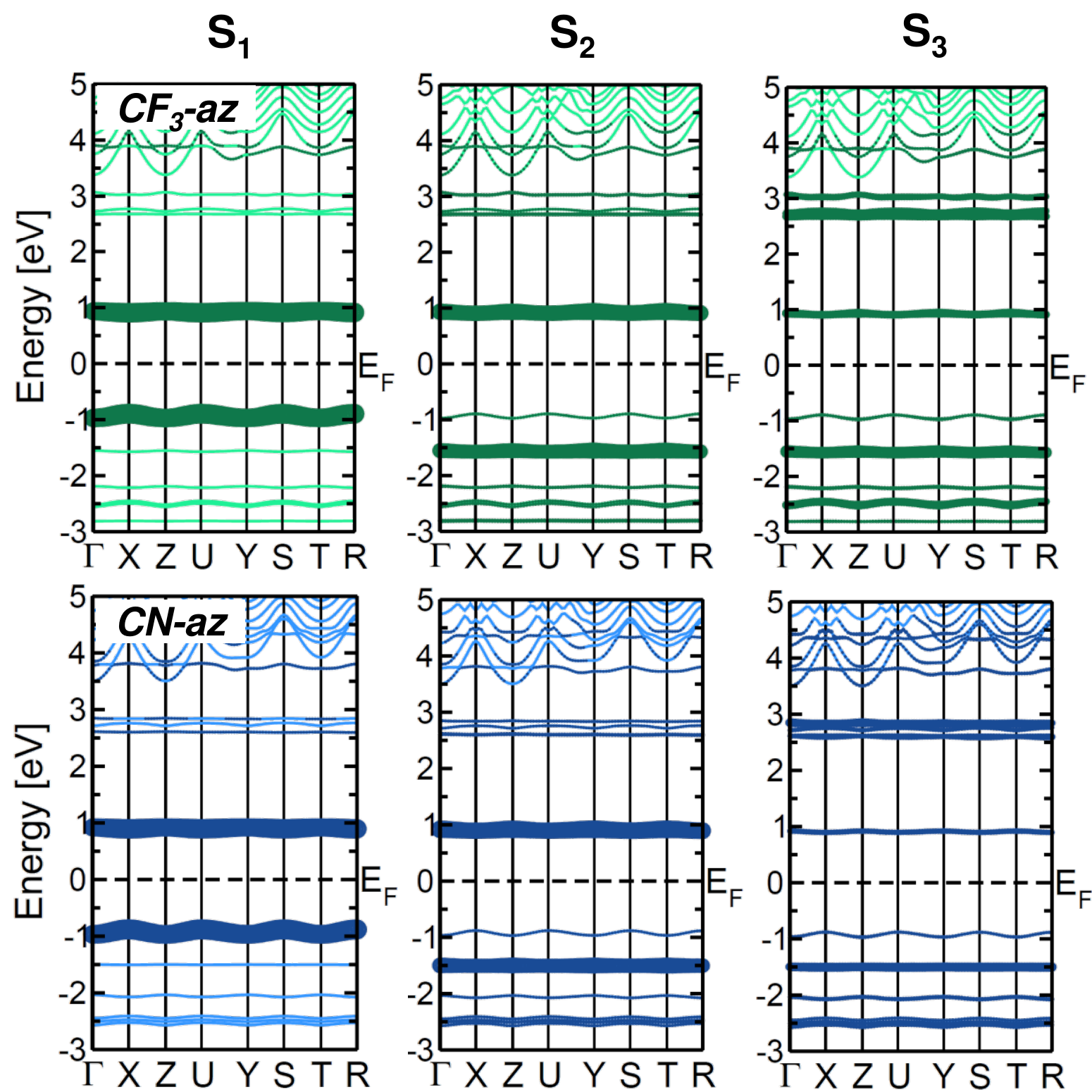}
\caption{(Color online) Band structure of \textit{diluted} SAMs. The Fermi energy ($E_F$) is set to zero in the mid-gap. Contributions from individual bands to the main excitons of the spectra are highlighted by colored circles of size representative of their magnitude.}
\label{fig:exciton_d-SAM}
\end{figure} 

To understand the effects of increasing packing density on the optical excitations of CF$_3$- and CN-functionalized azobenzene derivatives, we now turn to the diluted and packed SAMs.
The corresponding absorption spectra are shown in Fig.~\ref{fig:homo}.
In the case of the d-SAMs the main characteristics discussed above in the context of isolated molecules are still present, as a signature of the weak intermolecular interaction.
In particular, the dark S$_1$ excitation appears at lowest energy while the intense $\pi \rightarrow \pi^*$ resonance S$_2$ still dominates the near-UV region, followed by S$_3$ above 5 eV.
The effect of dense packing manifests itself in a reduced binding energy of the main excitations described above.
Except for the dark lowest-energy exciton S$_1$, the two peaks S$_2$ and S$_3$ are blue-shifted by approximately 0.5 eV compared to the gas-phase.
The closer distance between the chromphores enhances the screening and reduces the QP gap by about 200 meV.
As a consequence, especially for CN-az, the onset of the IPA spectrum is red-shifted compared to the one of the single molecule.

The character of the main excitations of the molecules is preserved in the d-SAM. 
The extremely low dispersion of the bands in the gap region (Fig.~\ref{fig:exciton_d-SAM}) indicates almost negligible interaction between these electronic states, which retain the character of the corresponding MOs in the gas-phase chromophore (Fig.~\ref{fig:KS}).
Hence, the lowest-energy excitation S$_1$ remains dipole-forbidden, with almost exclusive contribution from transitions between the valence-band maximum (VBM) and the conduction-band minimum (CBM).
The strong S$_2$ resonance dominates the near-UV region, stemming mainly from transitions between the VBM-1 and the CBM with additional contributions from deeper occupied bands and higher unoccupied ones.
The spectral region above 4.5 eV hosts a weaker peak, S$_3$, which exhibits a mixed character.
The single-particle transitions mostly contributing to this excitation involve $\pi$-like valence bands below the VBM and the five lowest conduction bands ($\pi^*$).
The exciton contributions spread over the entire BZ indicate the real-space localization of the excitons in the d-SAM and hence their molecular-like character.

\begin{figure*}
\center
\includegraphics[width=.8\textwidth]{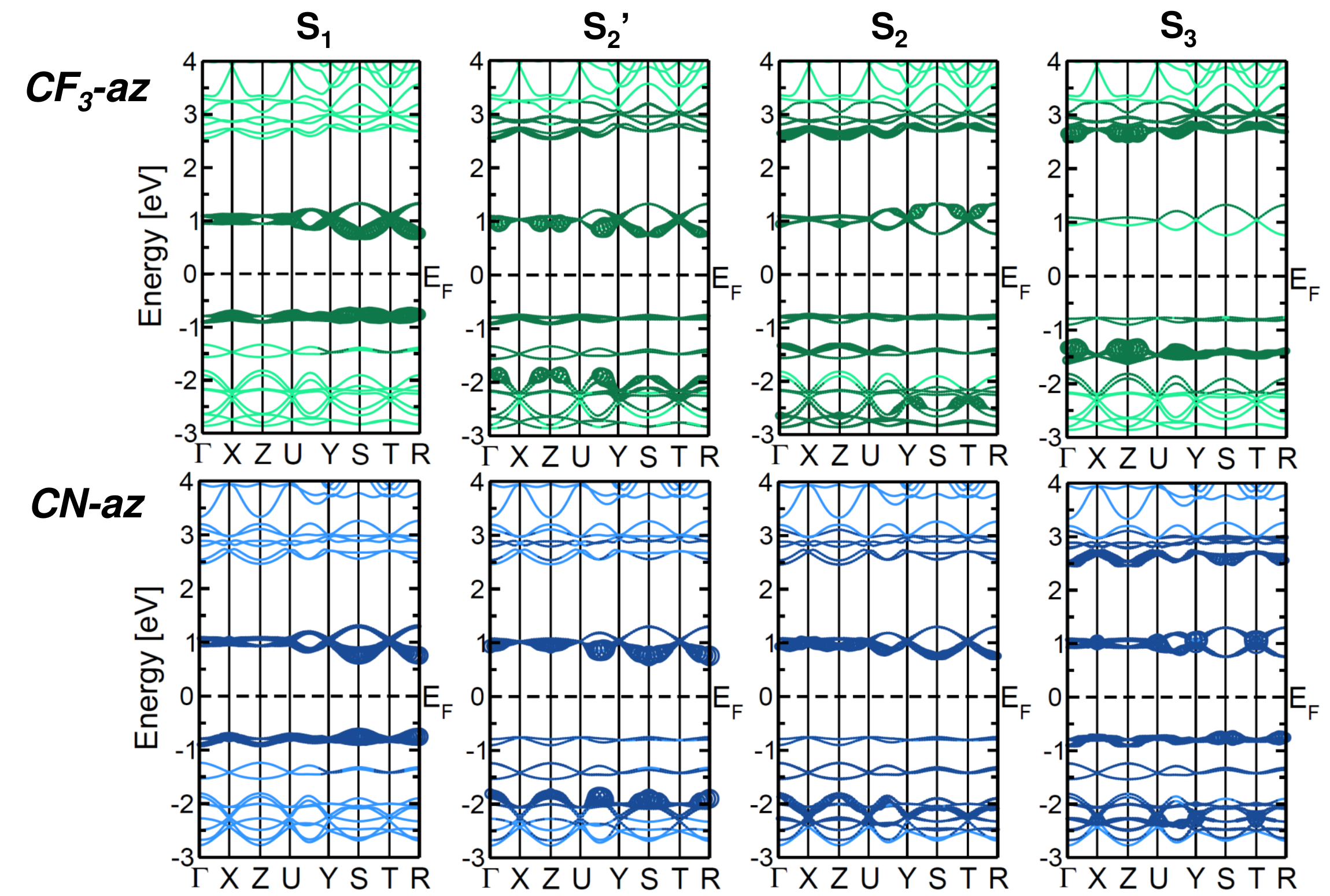}
\caption{(Color online) Band structure of \textit{packed} SAMs. The Fermi energy ($E_F$) is set to zero in the mid-gap. Contributions from individual bands to the main excitons of the spectra are highlighted by colored circles of size representative of their magnitude.}
\label{fig:exciton_p-SAM}
\end{figure*} 

The interaction between the chromophores gives rise to much more pronounced effects in the p-SAMs, where each molecule is  surrounded by six nearest-neighbors at a distance smaller than 4 \AA{}.
In this case, the spectral shape differs significantly from those of all the other systems considered so far.
The close aggregation of the chromophores leads to a pronounced intermolecular coupling, which is to a large extent driven by LFE, as extensively discussed in Ref.~\cite{cocc+16jcp}.
As a consequence, the spectral weight is significantly blue-shifted and the intense resonance in the near-UV tends to be smeared out, in agreement with the experimental findings for the CF$_3$-az SAM~\cite{gahl+10jacs}.
Although a more quantitative comparison with the measurements is obtained from the calculation of the differential reflectance~\cite{vorw+16cpc}, the energies and the intensities of the excitations are given already by Im$\epsilon_M$~\cite{cocc+16jcp}.
In the p-SAM, S$_1$ acquires finite (although very weak) oscillator strength, being activated by transition-dipole coupling between the chromophores.
Around 3.5 eV an additional peak appears, also due to transitions between the HOMO- and LUMO-derived bands.
Like S$_1$, these excitations are dipole-allowed only in the p-SAM, as an effect of the enhanced intermolecular coupling.
The broad band in the near-UV is formed by a large number of active excitations stemming from a manifold of $\pi \rightarrow \pi^*$ transitions. 
The binding energy of these excitations is drastically reduced compared to those in the d-SAMs and in the gas-phase systems.
This results from the interplay between the increased screening and the enhanced dipole coupling driven by the LFE and by the increased wave-function overlap~\cite{cocc+16jcp}.
Also the IPA onset is much broader in the p-SAM than in the other systems, due to the larger number of optically allowed vertical transitions.

\begin{figure*}
\center
\includegraphics[width=.8\textwidth]{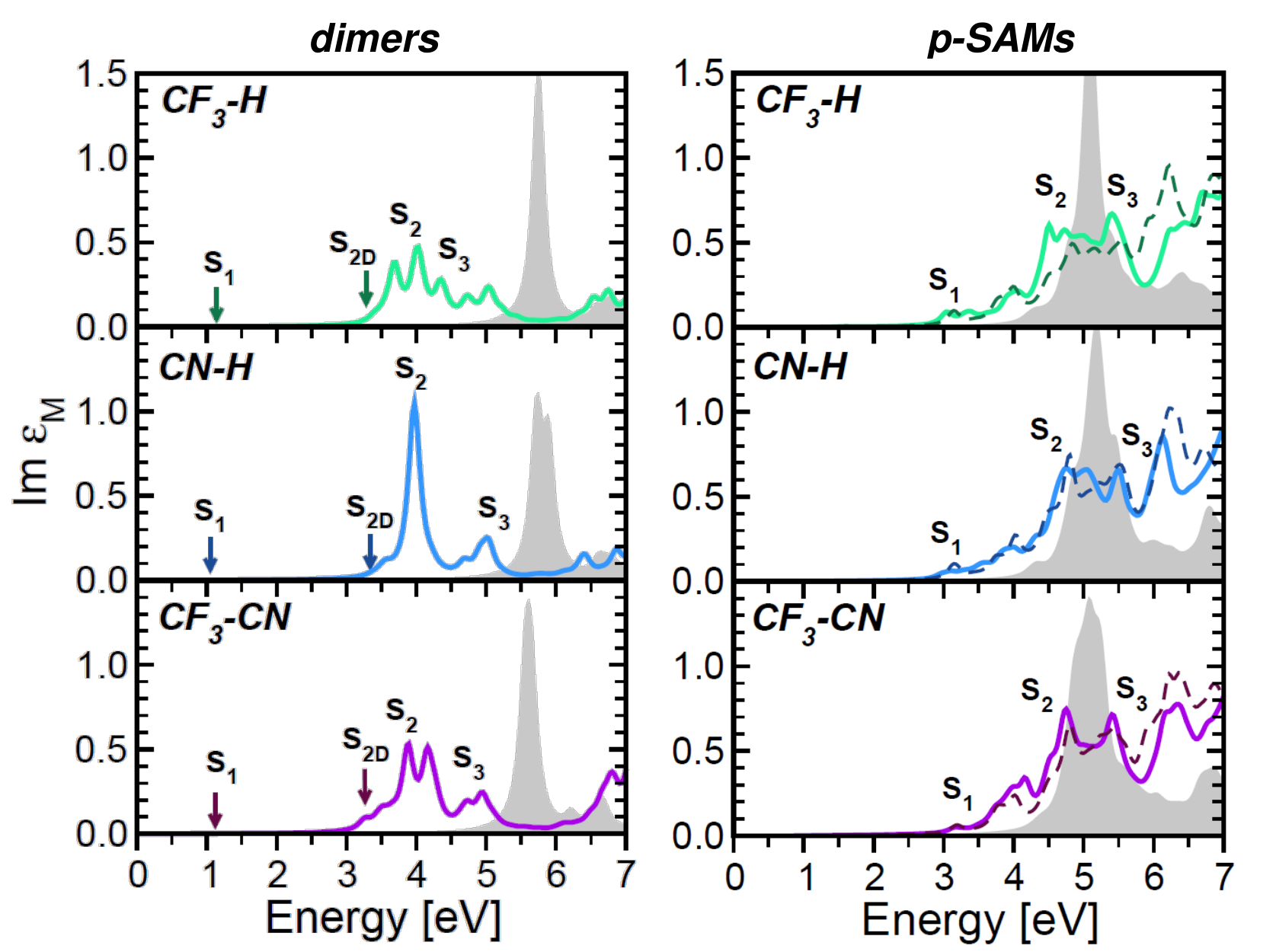}
\caption{(Color online) Optical absorption spectra of dimers (left) and p-SAMs (right) with mixed functionalizations, expressed by Im$\epsilon_M$ averaged over the three Cartesian components. Solid lines indicate the solutions of the BSE while filled gray areas correspond to the independent-particle approximation. Singlet excitons with vanishing oscillator strength are indicated by an arrow. For the p-SAMs, the averaged spectra of the homogenously functionalized SAMs with the corresponding terminations are also plotted (dashed lines). A Lorentzian broadening of 100 meV is applied to all spectra to mimic the excitation lifetime.}
\label{fig:mixed}
\end{figure*} 

By analyzing the band contributions to the main excitations (Fig.~\ref{fig:exciton_p-SAM}), we notice that the well-defined peak stemming from the HOMO-1 $\rightarrow$ LUMO transition in the isolated molecule does not appear here. 
Since each MO of the gas-phase chromophore is split into two sub-bands in the p-SAMs, due to the presence of two inequivalent molecules in the unit cell, the counterpart of such a resonance should arise from the VBM-2/VBM-3 to the CBM/CBM+1.
None of the excitations analyzed in Fig.~\ref{fig:exciton_p-SAM} presents clear contributions from these bands, regardless of the functionalization.
In the broad absorption band between 4 and 5 eV, we additionally identify S$_2$', another $\pi \rightarrow \pi^*$ excitation appearing at lower energy compared to S$_2$ representative of the manifold of excitations in the near-UV region~\cite{cocc+16jcp}.
In the CF$_3$-functionalized p-SAM a non-negligible contribution comes also from the HOMO-derived band (VBM/VBM-1).
In this system the S$_2$ excitation partially retains contributions from the VBM-2/VBM-3 to the CBM/CBM+1, but such transitions are strongly coupled to higher-energy ones.
In the CN-terminated p-SAM, instead, both S$_2$' and S$_2$ stem almost exclusively from deep valence bands targeting the CBM/CBM+1 and, especially in S$_2$, higher-energy unoccupied states.
The higher-energy excitation S$_3$ differs most significantly between the two p-SAMs (Fig.~\ref{fig:exciton_p-SAM}).
In the presence of CF$_3$-termination the involved single-particle transitions arise from the VBM-2 and VBM-3 to the manifold of unoccupied states above the LUMO-derived band.
In the CN-functionalized p-SAM a larger number of transitions contributes to the corresponding excitation.
All the conduction states up to the CBM+6 are involved, while in the valence region the HOMO-derived band and deeper states below the VBM-4 predominantly contribute.
We also notice that, except for S$_1$, whose band contributions are almost homogeneously spread in reciprocal space, the weight of higher-energy excitations tends to be more concentrated in specific regions of the BZ, pointing to an enhanced real-space delocalization of the $e$-$h$ pairs compared to the d-SAMs.

\subsection{Azobenzene-derived dimers and SAMs with mixed functionalizations}

\begin{table*}
\begin{tabular}{lccl}
\hline \hline
\multicolumn{4}{c}{\textit{CF$_3$-H}}\\
\hline 
$\#$ & Excitation energy & Binding energy & Contributions \\ \hline
S$_1$ & 1.14 & 3.79 & H-1 [H-az] $\rightarrow$ L+1 [H-az] (99$\%$)  \\ \hline
S$_{2D}$ & 3.28 & 1.65 & H-2 [H-az] $\rightarrow$ L [CF$_3$-az] (96$\%$)  \\ \hline
S$_2$ & 3.69 & 1.24 & H-2 [H-az] $\rightarrow$ L+1 [H-az] (56$\%$)  \\ \hline
S$_2$ & 3.98 & 0.95 & H-3 [CF$_3$-az] $\rightarrow$ L [CF$_3$-az] (71$\%$)  \\ \hline
\multirow{2}{*}{S$_3$}  & \multirow{2}{*}{4.33} & \multirow{2}{*}{0.60} & H-5 [H-az] $\rightarrow$ L+1 [H-az] (27$\%$)  \\ 
& & & H-5 [H-az] $\rightarrow$ L [CF$_3$-az] (40$\%$) \\ 
\hline \hline
\multicolumn{4}{c}{\textit{CN-H}}\\ 
\hline
$\#$ & Excitation energy & Binding energy & Contributions \\ \hline
S$_1$ & 1.05 & 3.79 & H-1 [CN-az] $\rightarrow$ L [CN-az] (96$\%$)  \\ \hline
S$_{2D}$ & 3.34 & 1.50 & H-2 [H-az] $\rightarrow$ L [CN-az] (96$\%$)  \\ \hline
\multirow{2}{*}{S$_2$} & \multirow{2}{*}{3.97} & \multirow{2}{*}{0.87} & H-2 [H-az] $\rightarrow$ L+1 [H-az] (52$\%$) \\ 
& & & H-3 [CN-az] $\rightarrow$ L [CN-az] (31$\%$) \\ \hline
\multirow{2}{*}{S$_3$}  & \multirow{2}{*}{4.69} & \multirow{2}{*}{0.15} & H-3 [CN-az] $\rightarrow$ L+4 [CN-az] (21$\%$)  \\ 
& & & H-3 [CN-az] $\rightarrow$ L+1 [H-az] (15$\%$) \\ 
\hline \hline
\multicolumn{4}{c}{\textit{CF$_3$-CN}}\\ 
\hline
$\#$ & Excitation energy & Binding energy & Contributions \\ \hline
S$_1$ & 1.12 & 3.82 & H-1 [CN-az] $\rightarrow$ L [CN-az] (90$\%$)  \\ \hline
\multirow{3}{*}{S$_{2D}$}  & \multirow{3}{*}{3.26} & \multirow{3}{*}{1.68} & H-3 [CN-az] $\rightarrow$ L [CN-az] (47$\%$)  \\ 
& & & H-4 [CN-az] $\rightarrow$ L [CN-az] (26$\%$) \\ 
& & & H-2 [CF$_3$-az]  $\rightarrow$ L [CN-az] (11$\%$) \\ \hline
\multirow{2}{*}{S$_{2}$}  & \multirow{2}{*}{3.88} & \multirow{2}{*}{1.06} & H-3 [CN-az] $\rightarrow$ L [CN-az]  (25$\%$)  \\ 
& & & H-4 [CN-az]  $\rightarrow$ L [CN-az]  (26$\%$) \\ \hline
S$_{2}$ & 4.16 & 0.79 & H-7 [CF$_3$-az] $\rightarrow$ L+1 [CF$_3$-az] (12$\%$)  \\ \hline
\multirow{2}{*}{S$_3$}  & \multirow{2}{*}{4.73} & \multirow{2}{*}{0.21} & H-2 [CF$_3$-az] $\rightarrow$ L+4 [CF$_3$-az] (22$\%$)  \\ 
& & & H-7 [CF$_3$-az] $\rightarrow$ L+1 [CF$_3$-az] (15$\%$) \\ 
\hline
\hline
\end{tabular}
\caption{Excitation energies, binding energies, and dominant single-particle contributions (weight in parenthesis) of the excitations of the dimers with mixed functionalization marked in Fig.~\ref{fig:mixed}. The predominant character of the MOs is reported in squared brackets. All energies are in eV.}
\label{tab:mix-dimers}
\end{table*}

To investigate the effects of mixing different functional groups within the same sample, we restrict our attention to the p-SAMs and to their non-periodic counterparts, \textit{i.e.} the dimers, from which we start our analysis.
While the excitations highlighted and discussed in Sec.~\ref{sec:homogeneous} can be still identified in Fig~\ref{fig:mixed}, the overall spectral shape and the intensity of the peaks are quite peculiar, especially in the CF$_3$-H and the CF$_3$-CN dimers.
In the near-UV region the spectrum of these systems is no longer dominated by the sharp S$_2$ resonance but by two peaks of approximately half its intensity.
To elucidate the origin of this behavior, we consider the composition of the excitations reported in Table~\ref{tab:mix-dimers}.
As discussed for the homogeneously functionalized systems in Sec.~\ref{sec:homogeneous}, each MO of the molecules generates two electronic levels in the dimers.
Although the spatial distribution of the wave-function depends on the termination of the azobenzene molecules, the $\pi$/$\pi^*$ orbitals of homogeneously functionalized dimers tend to spread over both chromophores, as a consequence of intermolecular coupling (Fig.~\ref{fig:KS}).
This is not the case of the mixed dimers, where different adjacent groups reduce the interaction between the chromophores, thereby partially inhibiting wave-function hybridization (see Table~\ref{tab:mix-dimers}, where the predominant localization of each MO is indicated in square brackets).
As a result, two separate bright resonances with $\pi \rightarrow \pi^*$ character originate from the transitions between the HOMO--1-like and the LUMO-like states of each monomer.
Hence, for the CF$_3$-H and the CF$_3$-CN dimers two S$_2$ excitations are listed in Table~\ref{tab:mix-dimers}.
The energy separation between them is approximately 300 meV in the BSE spectra, while it decreases to less than 100 meV in the IPA ones, giving rise to only one peak in Im$\epsilon_M$.
On the contrary, in the CN-H dimer the single-particle transitions involving the $\pi$/$\pi^*$ orbitals localized on each molecule are separated by more than 150 meV, resulting in two maxima in the IPA spectrum.
However, these transitions couple when excitonic effects are accounted for, thereby generating the intense S$_2$ resonance in the BSE spectrum (Fig.~\ref{fig:mixed}).

\begin{figure}
\center
\includegraphics[width=.5\textwidth]{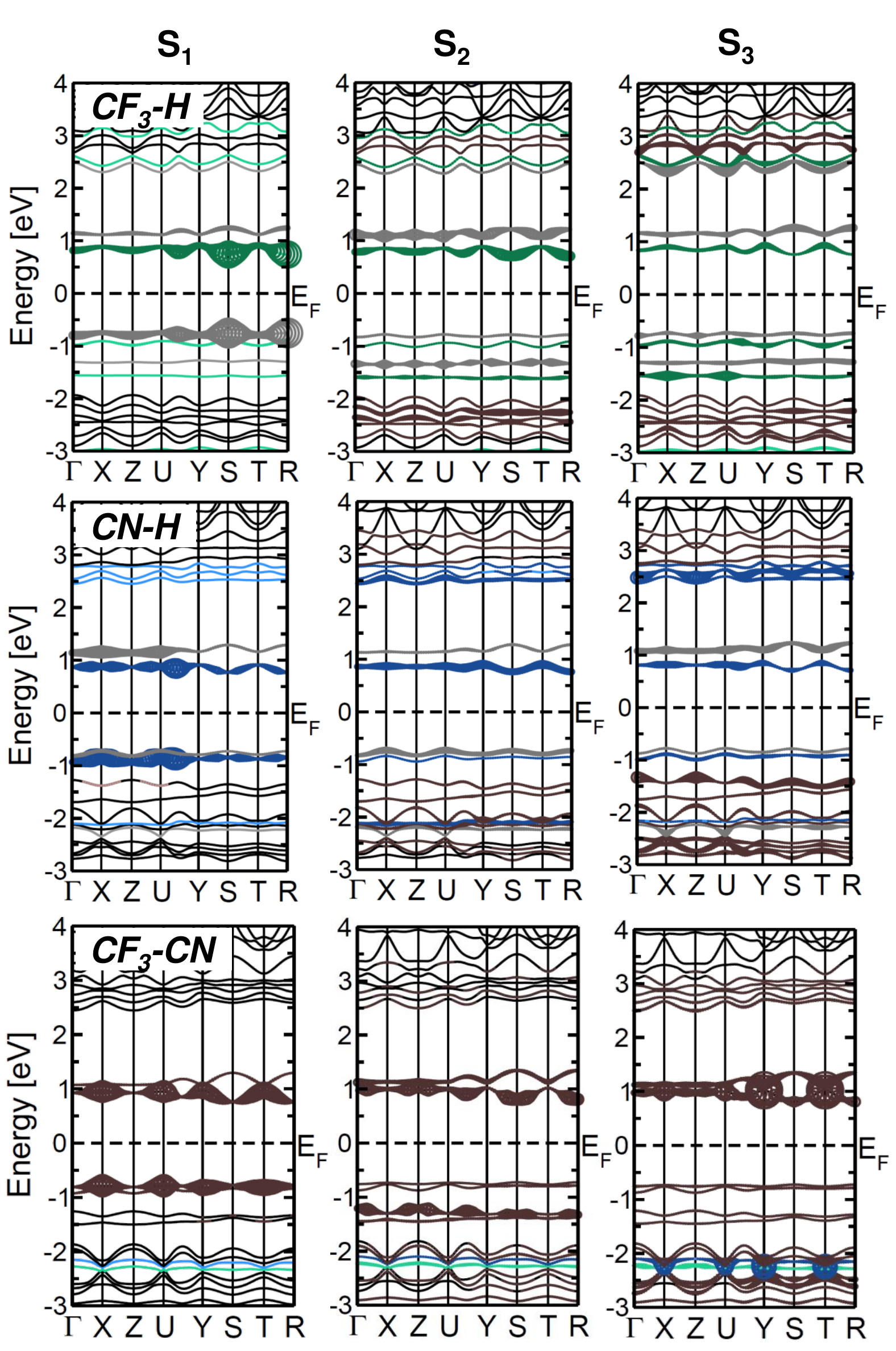}
\caption{(Color online) Band structure of p-SAMs with mixed functionalization. The Fermi energy ($E_F$) is set to zero in the mid-gap. Contributions from individual bands to the main excitons of the spectra are highlighted by colored circles of size representative of their magnitude. Green, blue, and gray circles indicate bands with predominant contribution from the CF$_3$-, CN-, and H-terminated molecule, respectively. Brown circles are adopted for hybridized bands, where the single-particle states are uniformly distributed on both molecules in the unit cell.}
\label{fig:excitons-mixed}
\end{figure} 

In all dimers we can identify the weak Davydov component at lower energy compared to S$_2$.
Notably, in these systems S$_{2D}$ stems from transitions between occupied and virtual MOs localized on different molecules. 
Hence, the low intensity of this excitation is not only determined by the cancellation of transition-dipole components, but also by minimized wave-function overlap.
In the CF$_3$-CN dimer S$_{2D}$ has a more mixed character, with relevant contributions from $\pi \rightarrow \pi^*$ transitions between MOs localized on the CN-functionalized chromophore.
As a consequence, this excitation gains oscillator strength compared to its counterparts in the other two dimers, while the other excitations are less significantly affected by mixed functionalizations.
S$_1$ remains dipole-forbidden as in the isolated molecule and S$_3$ retains its mixed character (Table~\ref{tab:mix-dimers}).
The excitation energies are not significantly affected by the presence of azobenzene derivatives with different terminations. 
The absorption onsets of the spectra of the three dimers are almost aligned with respect to each other, and also the exciton binding energies are very similar.

Finally, we turn to the p-SAMs with mixed functionalizations. 
The resulting absorption spectra are displayed in the right panel of Fig.~\ref{fig:mixed}.
In light of the differences between the homogeneously functionalized and the mixed dimers discussed above, it is instructive to compare the spectra of the mixed p-SAMs with those of their counterparts including a single functionalization.
To do so, we plot on top of the Im$\epsilon_M$ of the mixed SAMs the same quantity obtained by averaging the contributions of the homogeneously functionalized SAMs with the corresponding terminations.
From this analysis it is apparent that the presence of different terminations in the same SAM does not change the main characteristics of the optical absorption. 
This is consistent with the behavior of the dimers, where the wave-function hybridization between the chromophore is partially inhibited by the coexistence of different functionalizations.
The localization of the electronic states is highlighted in Fig.~\ref{fig:excitons-mixed}.
KS states with predominant distribution on the H-passivated azobenzene derivative are marked in gray, while those mostly on CF$_3$- and CN-terminated molecules are colored in green and blue, respectively.
Hybridized bands, where the electronic wave-function is spread over both inequivalent molecules, are brown.

We consider first the two mixed SAMs containing H-passivated chromophores, namely CF$_3$-H and CN-H.
In both cases the two highest valence bands, that stem from the $n$-like HOMO of the molecule and the two lowest conduction bands, corresponding to the $\pi^*$ LUMO of the gas-phase azobenzene derivative, are more significantly separated in energy compared to their homogeneously functionalized counterparts (see Fig.~\ref{fig:exciton_p-SAM} and Ref.~\cite{cocc+16jcp}).
This is particularly evident in the conduction region, where the CBM and the CBM+1 never cross. 
The same behavior occurs also for the $\pi$ ($\pi^*$) valence (conduction) states VBM-2 and VBM-3 (CBM+2 and CBM+3) of the CF$_3$-H and CN-H SAMs, although in the latter the $\pi$ states VBM-2 and VBM-3 are delocalized on both molecules.
Further away from the gap, hybridization effects become more prominent, and the bands of both systems are more delocalized.
The situation is different for the CF$_3$-CN p-SAM, where almost all electronic states displayed in Fig.~\ref{fig:excitons-mixed} (bottom panel), are spread between the two inequivalent molecules in the unit cell.
This behavior suggests that the combination of two relatively large functional groups, such as $-$CF$_3$ and $-$CN, promotes intermolecular coupling already at a single-particle level.
Remarkably, the spectra in Fig.~\ref{fig:mixed} given by the averaged contributions of the homogeneously functionalized SAMs are very similar to those computed for the mixed ones.
This coincidence suggests that the chromophores with a given functionalization interact weakly with the nearest neighbors carrying a different termination.
Like in the homogeneously functionalized p-SAMs, S$_1$ is no longer dipole-forbidden, although significantly weaker than the local maxima of Im$\epsilon_M$ in the near-UV region.
However, both S$_2$ and S$_3$ give rise to much better defined peaks than in their homogeneously functionalized counterparts (see Fig.~\ref{fig:homo}).
In addition, at the onset of the IPA spectra of the CF$_3$-az and CN-az SAMs a strong resonance appears instead of the rather featureless hump.
We ascribe this behavior to the reduced band hybridization induced by the presence of different functional groups.
More pronounced differences between the spectra of the mixed SAMs and the averaged result of the homogeneously functionalized ones are visible at higher energies.
In this region the excitations have a more mixed character~\cite{cocc+16jcp}, with a large number of interband transitions involved in the formation of the $e$-$h$ pairs.
The latter are therefore more significantly influenced by the specific functionalization of the azobenzene derivatives compared to the excitons in the near-UV region, which are driven solely by the KS states corresponding to the HOMO- and LUMO-derived bands in the gas-phase compound.

\section{Summary and conclusions}
In summary, we have performed a detailed analysis on the optical properties of azobenzene-functionalized SAMs, addressing the interplay between packing density of the chromophores and their functionalization with the $-$CF$_3$ and the $-$CN marker groups.
We have shown that the presence of functional groups does not significantly affect the spectral features, which are instead determined by the electronic properties of the azobenzene backbone and by the packing of the chromophores. 
While the main absorption band in the near-UV region, which is known to trigger the photo-isomerization process, dominates the spectra of the systems with low molecular density, it is suppressed in the closely-packed SAMs.
The enhanced dipole coupling in the excited states blue-shifts the spectral weight and is responsible for a significant broadening of the absorption bands.
This finding also suggests that the correct spectral features of dense SAMs cannot be reproduced by isolated dimers or small clusters, but are captured only in a periodic system.

Mixing different functionalizations partially inhibits the coupling between the chromophores, which thereby preserve molecular-like features.
While separate $\pi$-$\pi^*$ resonances appear in the spectra of the dimers, especially when the CF$_3$- and CN-functionalized azobenzene derivatives are mixed with the H-passivated ones, in the packed SAMs the blue-shift and the broadening of the main absorption band is not totally quenched.
Hence, the presence of mixed functionalization partially counteracts the intermolecular coupling promoted by the high molecular density.
As an outlook, we expect that accurate molecular-dynamical simulations, already applied successfully to azobenzene SAMs on gold substrates~\cite{pipo+13lang}, would yield a more realistic description of the mixed SAMs and thereby provide a more accurate starting point for the analysis of their electronic and optical properties.
To conclude, our results suggest that a tailored interplay between steric separation of the chromophores in the SAM and their functionalization can be adopted to design well-ordered molecular architectures with improved photo-switching capabilities.

Input and output files of the calculations are available free of charge in the NOMAD Repository (http://nomad-repository.eu/) at the following link: http://dx.doi.org/10.17172/NOMAD/2017.06.30-1
\subsection*{Acknowledgments}
We thank C. Gahl and P. Saalfrank for stimulating discussions.
This work was funded by the German Research Foundation, through the Collaborative Research Center 658, ``Elementary Processes in Molecular Switches at Surfaces". C.C. acknowledges additional financial support from the Berliner Chancengleichheitsprogramm (BCP) and from IRIS Adlershof.

\end{document}